\documentclass[preprint,10pt,twocolumn, 3p]{elsarticle}

\usepackage{lineno,hyperref}
\modulolinenumbers[5]

\newcounter{bla}

\journal{Optics Communications}

\usepackage{amsfonts}
\expandafter\let\csname equation*\endcsname\relax
\expandafter\let\csname endequation*\endcsname\relax
\usepackage{amsmath}
\usepackage{booktabs}
\usepackage{tabularx}
\usepackage{wasysym}
\usepackage{mathtools}
\usepackage{siunitx}
\usepackage{nicefrac}
\DeclarePairedDelimiter\abs{\lvert}{\rvert}%
\usepackage{graphicx}
\newcommand{\CC}{%
    {\settoheight{\dimen0}{C}C\kern-.05em \resizebox{!}{\dimen0}{\raisebox{\depth}{++}}}}
\newcommand{\CCC}{%
    {\settoheight{\dimen0}{C}C/C\kern-.05em \resizebox{!}{\dimen0}{\raisebox{\depth}{++}}}}
\newcommand{\CS}{%
    {\settoheight{\dimen0}{C}C\kern-.05em \resizebox{!}{\dimen0}{\raisebox{\depth}{\#}}}}
\graphicspath{{Figs/}}

\begin{document}
    \suppressfloats 
    
    \begin{frontmatter}
        
        \title{Relative limitations of increasing the number of modulation levels in computer generated holography}
        
        \author[mymainaddress]{Peter J. Christopher\corref{mycorrespondingauthor}}
        \cortext[mycorrespondingauthor]{Corresponding author}
        \ead{pjc209@cam.ac.uk}
        \ead[url]{www.peterjchristopher.me.uk}
        
        \author[mymainaddress]{Timothy D. Wilkinson}
        
        \address[mymainaddress]{Centre of Molecular Materials, Photonics and Electronics, University of Cambridge}
        
        \begin{abstract}
            Phase and amplitude spatial light modulators (SLMs) capable of both binary and multi-level modulation are widely available and offer a wide range of technologies to choose from for holographic applications. While the replay fields generated with multi-level phase-only SLMs are of a significantly higher quality than those generated by equivalent binary phase-only SLMs, evidence is presented in this letter that this improvement is not as marked for amplitude SLMs, where multi-level devices offer only a small benefit over their binary counterparts. Heuristic and numerical justifications for this are discussed and conclusions drawn.
        \end{abstract}
        
        
    \end{frontmatter}
    
    \section{Introduction}
      
	Holograms exploit the diffractive nature of light to project a desired light field or intensity pattern onto a given surface. While significant attention has been given to using holograms for display purposes, the underpinning technology has also enabled other technologies such as lithography~\cite{Turberfield2000} and optical tweezing~\cite{Melville2003,Grieve2009}. Holograms were first created by exposing photographic plates in the 1960s~\cite{gabor1948new, Leith:62}, with the field of holography seeing renewed interest in the 1980s due to the introduction of the computer-controlled spatial light modulator (SLM) that allow the spatial profile of an incident beam to be modulated. This was in conjunction to an increase in the available numerical processing power.
		
	Most SLMs are capable of modulating either the amplitude or the phase in exclusion of the other. While systems are available that allow both amplitude and phase to be modulated in a coupled manner \cite{Jesacher:08}, it is currently challenging to modulate both regimes of light directly, and this is normally achieved by incorporating phase-only or magnitude-only SLMs into larger optical set-ups~\cite{Shibukawa:14}. Furthermore, the digital nature of these devices entails that the phase or amplitude levels available are typically discretised. Nematic liquid crystal on silicon devices typically offer 256 phase levels, for example, and are capable of switching at speeds of over $50~Hz$. Alternatively, ferroelectric liquid crystal on silicon (LCoS) and digital micromirror devices (DMD) offer switching speeds of the order of $kHz$, but are only capable of binary phase or amplitude modulation. Table \ref{fig:LCTypes} provides an overview of some commonly used liquid crystal technologies.
    
    An appropriate choice of SLM for the application to hand is hence important, and the restrictions imposed by the SLM have to be accounted for when generating a suitable hologram for display. A clear advantage of binary SLMs is their switching speed, but for static holograms a user might expect a multi-level SLM to impose fewer constraints and to yield a field pattern closer to the desired target image. This paper aims to show that, while this intuition is true in the case of multi-phase SLMs, this is not the case for multi-amplitude SLMs that are found to not offer a marked improvement over binary-amplitude SLMs. Numerical and heuristic justifications of this are also presented.
		
	\begin{table}[tbhp]
        \caption{Modulation achieved by different liquid crystal on silicon SLMs \cite{de1997complex}.   }
        \begin{tabular}{m{1.1cm} m{2.2cm} | m{1.1cm} m{2.2cm}}
            \hline\noalign{\smallskip}
            \multicolumn{1}{c}{LC}                       & \multicolumn{1}{c}{Complex}                                                         & \multicolumn{1}{c}{LC}                       & \multicolumn{1}{c}{Complex}                                                         \\
            \multicolumn{1}{c}{Type}                     &  \multicolumn{1}{c}{Modulation}                                                      & \multicolumn{1}{c}{Type}                     &  \multicolumn{1}{c}{Modulation}                                                      \\
            \noalign{\smallskip}\hline\noalign{\smallskip}
            Nematic (Parallel Polarisers)                 & \includegraphics[trim={0 0 0 0},width=1.0\linewidth,page=1]{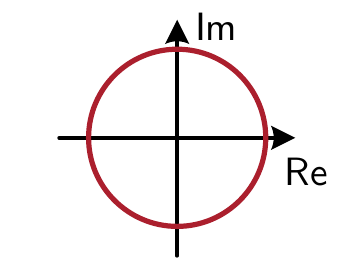} & Nematic (Perpendicular Polarisers)                  & \includegraphics[trim={0 0 0 0},width=1.0\linewidth,page=2]{liquidcrystaltypes.pdf} \\[2pt]
            Twisted \newline Nematic & \includegraphics[trim={0 0 0 0},width=1.0\linewidth,page=3]{liquidcrystaltypes.pdf} & Twisted \newline Nematic & \includegraphics[trim={0 0 0 0},width=1.0\linewidth,page=4]{liquidcrystaltypes.pdf} \\[2pt]
            Smectic A                & \includegraphics[trim={0 0 0 0},width=1.0\linewidth,page=5]{liquidcrystaltypes.pdf} & Smectic B                & \includegraphics[trim={0 0 0 0},width=1.0\linewidth,page=6]{liquidcrystaltypes.pdf} \\[2pt]
            Antiferro- electric      & \includegraphics[trim={0 0 0 0},width=1.0\linewidth,page=7]{liquidcrystaltypes.pdf} & Twisted  SmC             & \includegraphics[trim={0 0 0 0},width=1.0\linewidth,page=8]{liquidcrystaltypes.pdf} \\[2pt]
            \noalign{\smallskip}\hline
        \end{tabular}
        \label{fig:LCTypes}
    \end{table}
        
    \section{Background}
    
    Consider an SLM displaying a complex-valued hologram $f(x,y)$ and illuminated by a plane wave. The complex-valued two-dimensional light field pattern $F(u,v)$ created in the far-field, also known as the replay field, is the Fourier transform of the illuminated hologram~\cite{goodman2005introduction}. Assuming regular sampling of both the hologram plane and the far field plane, the transform between the two planes can be efficiently calculated using a discrete Fourier transform (DFT) and its inverse as per Eq. \ref{fouriertrans2d5c} where $x$ and $y$ are spatial coordinates in the hologram field, $u$ and $v$ are spatial coordinates in the far field and both fields are discretised into $N_xN_y$ points.
    
    \begin{align}
        F_{u,v}      & = \frac{1}{\sqrt{N_xN_y}}\sum_{x=0}^{N_x-1}\sum_{y=0}^{N_y-1} f_{xy}e^{-2\pi i \left(\frac{u x}{N_x} + \frac{v y}{N_y}\right)} \label{fouriertrans2d5c}   \nonumber\\
        f_{x,y} & = \frac{1}{\sqrt{N_xN_y}}\sum_{u=0}^{N_x-1}\sum_{v=0}^{N_y-1} F_{uv}e^{2\pi i \left(\frac{u x}{N_x} + \frac{v y}{N_y}\right)} 
    \end{align}

	Practically, this means that a hologram needs to be generated such that its Fourier transform yields the desired far-field pattern. Given an SLM where the amplitude and phase of each pixel can be controlled independently and in a continuous fashion, this would correspond to merely taking the inverse DFT of the desired far-field pattern. In a practical setting, the constraints of the SLM also need to be taken into account, opening up an entire field of hologram generation algorithms and techniques.
		
	Some applications of holography, such as holographic displays, only require the intensity of the obtained replay field to match the target replay field. In this case the phase insensitive mean-squared error (MSE) of Eq.~\ref{eq:msepi} can be used as a metric to compare and optimise holograms. Other applications, such as the holographic control of optical fibre modes \cite{fontaine2019laguerre}, require both the magnitude and phase of the obtained and target replay fields to be compared.
	
	\begin{equation} \label{eq:msepi}
	   E_{\text{MSE}}(T,R) = \frac{1}{N_x N_y}\sum_{N_x}\sum_{N_y} \left[\abs{T_{u,v}} -  \abs{R_{u,v}}\right]^2.
	\end{equation} 
    
    where $T$ and $R$ represent the target and observed replay fields respectively. In this work we only consider the phase insensitive case though we believe the results are likely to be equally applicable to the phase sensitive case.
    
    \section{Test Image}
    
\begin{figure}[tbhp]
    	\centering
    	{\includegraphics[trim={0 0 0 0},width=0.29\linewidth,page=1]{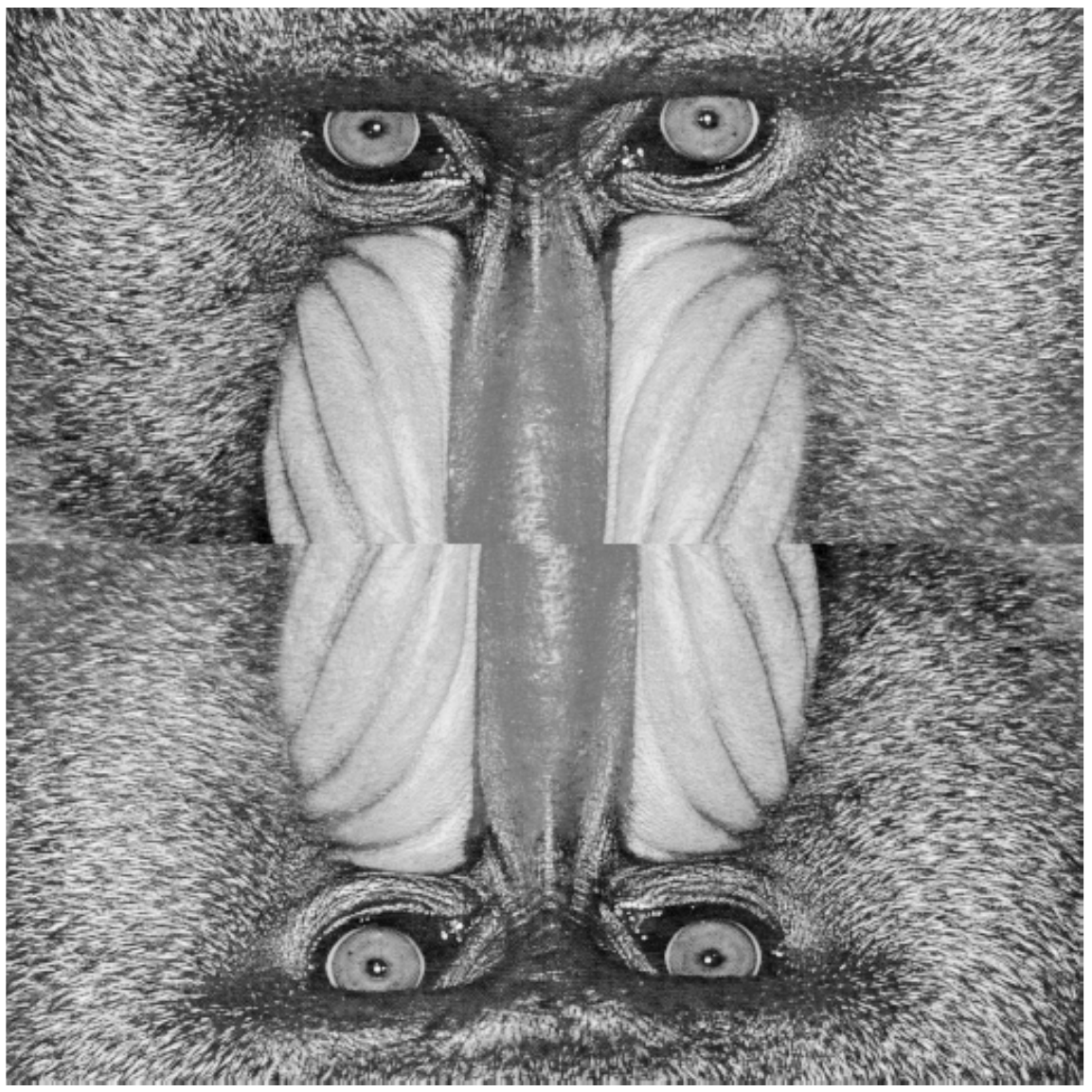}}
        {\includegraphics[trim={0 0 0 0},width=0.69\linewidth,page=1]{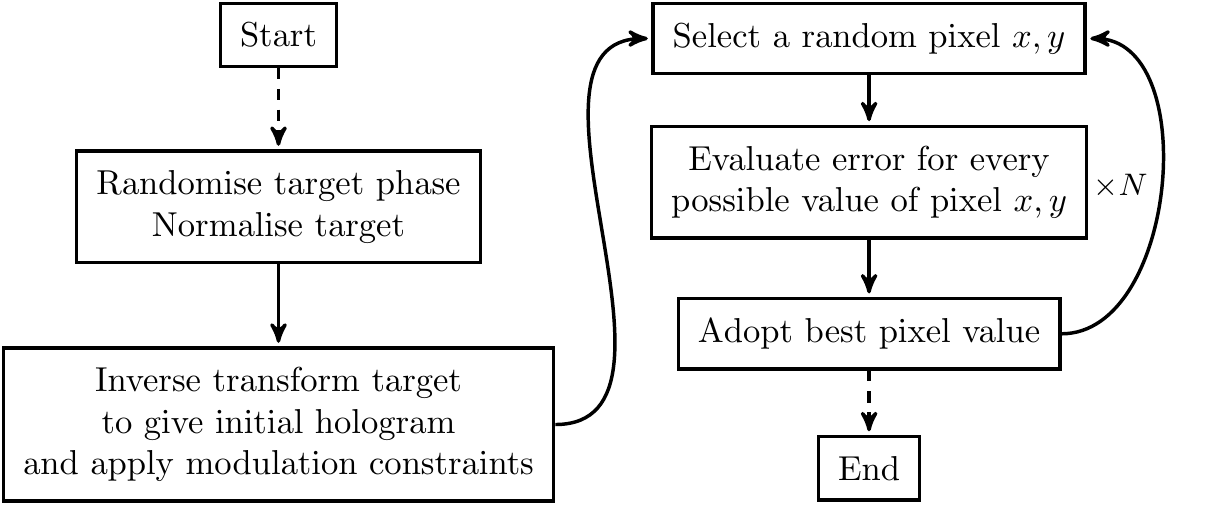}}
    	\caption{Mandrill test image magnitudes with artificial rotational symmetry (left) and flow chart detailing the hologram generation algorithm used (right)}
    	\label{fig:MandrillRotSym}
    \end{figure}

    The test image used for this work is the \textit{Mandrill} image from the USC-SIPI database~\cite{sipidatabaseref} with a randomised phase profile. The conjugate symmetry of the Fourier transform necessitates that all on-axis amplitude holograms have $180^{\circ}$ symmetry. To improve the comparison between phase and amplitude holograms, we artificially add rotational symmetry to the Mandrill test image as shown in Figure~\ref{fig:MandrillRotSym}~(left).
    
	\section{Single Pixel Modification} \label{single}
    
    In order to discuss the impact of additional modulation levels on the hologram reconstruction we first discuss the effect of changing a single pixel independently.
    
\begin{figure}[tbhp]
    	\centering
    	{\includegraphics[trim={0 0 0 0},width=0.47\linewidth,page=1]{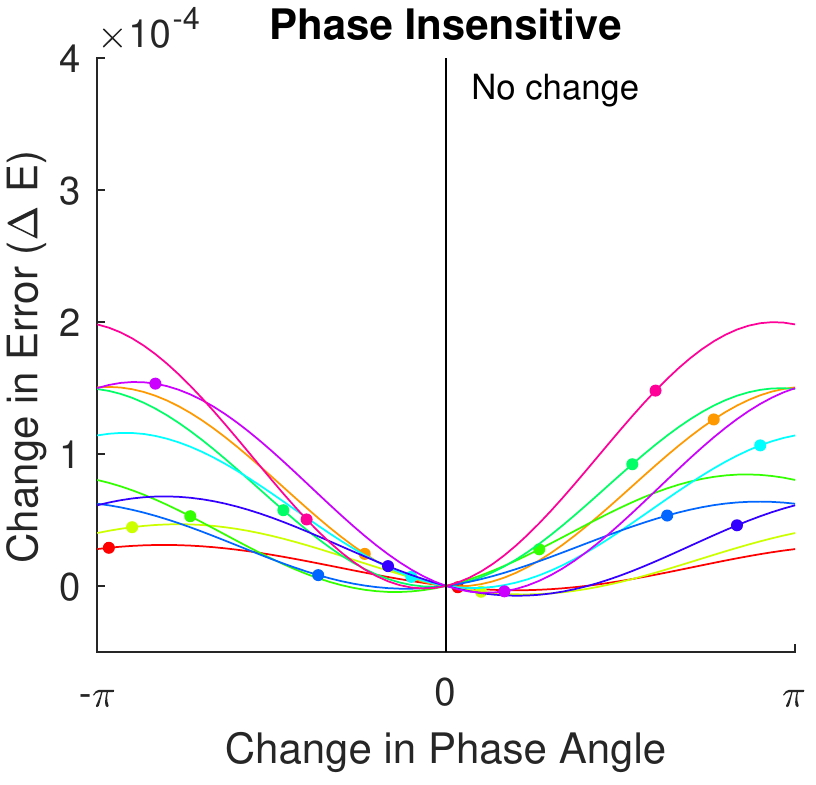}}
        {\includegraphics[trim={0 0 0 0},width=0.49\linewidth,page=1]{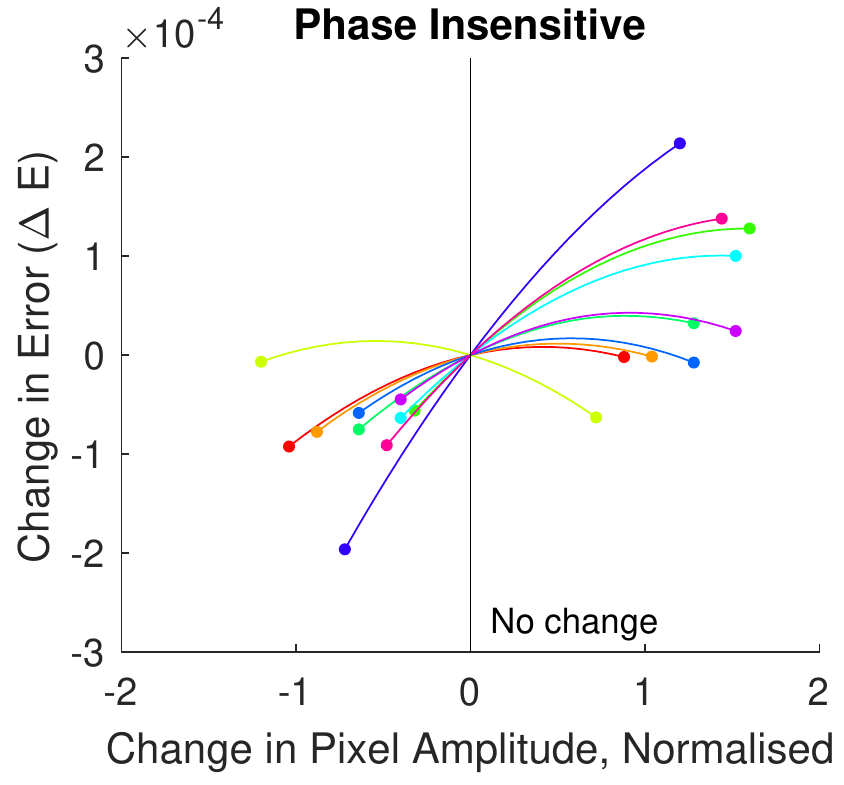}}
    	\caption{Effect of changing the value of a random selection of ten individual pixels on the MSE for phase (left) and amplitude (right) holograms. Test image used is a 512 $\times$ 512 pixel version of Mandrill with induced rotational symmetry as shown in Figure~\ref{fig:MandrillRotSym}(right). Dots the values achievable on a binary device.}
    	\label{fig:changes}
    \end{figure}

    The inverse DFT of the Mandrill image of Fig.~\ref{fig:MandrillRotSym} (left) is taken and the magnitude of all obtained pixel values normalised in accordance with Parseval's theorem, corresponding to the initial modulation required for a phase-only SLM. The phase of a single randomly selected hologram pixel is altered by a random angle in the range $[-\pi,\pi]$ relative to this initial phase angle, and the resultant error $\Delta E$ is calculated for the replay field. Ten independent tests are shown in Fig.~\ref{fig:changes} (left). The plotted lines shown represent the errors achievable on a multi-level device for a given phase angle while the dots show the errors shown achievable on a binary device. It is observed that the binary phase modulation values rarely correspond to the lowest error achievable on a multi-level device. For our test image, a $2^8$ multi-level phase SLM would offer a lower MSE when compared to a binary phase modulator more than $99\%$ of the time.
    
    To investigate modulation of an amplitude-only SLM the inverse DFT of the Mandrill image of Fig. \ref{fig:MandrillRotSym} is again taken, but in this case the phase-angle of each pixel is set to $0^{\circ}$. The amplitude of ten randomly selected pixels is then varied in the range $\left(0,2\right)$ with the associated change in error $\Delta E$ calculated with results shown in Fig. \ref{fig:changes} (right). In this case, it can be seen that the lowest error value often occurs at extremal values, corresponding to the values that can be obtained with an equivalent binary SLM. For our test image, a $2^8$ amplitude level SLM would have offered better error improvements over a binary amplitude modulator less than $3\%$ of the time.
     
    The combination of these two results suggests that multi-phase modulation offers improvements over binary-phase modulation for almost all pixels individually while multi-amplitude modulation does not offer similar improvements over binary-amplitude modulation. To explore this further, we consider the case of generating an entire hologram.	
    
    \section{Algorithms}
    
    A wide array of hologram generation algorithms are available but many of these are limited to a specific class of problem. For example, Gerchberg Saxton (GS) is commonly applied only to phase modulated holography \cite{gerchberg1972practical}. In order to provide the fairest comparison of binary  vs. multi-level devices and amplitude modulating vs. phase modulating devices we adopt a form of two common search algorithms: direct search (DS) \cite{DBS_Original_1, seldowitz1987synthesis} and simulated annealing (SA) \cite{kirkpatrick1983optimization}. Both algorithms take the inverse DFT  of the target complex replay field to obtain the associated initial hologram. This hologram is then \textit{quantised} to conform to the modulation constraints of the SLM, normally done by changing the value of each pixel to the nearest permitted value. The various algorithms then take different approaches to changing pixels within the modulation constraints in order to reduce the error of the new replay field. 
    
    We take a lightly modified algorithm, that shown in Fig.~\ref{fig:MandrillRotSym} (right). This is similar to a direct search algorithm, except that the pixel under consideration is set to the best possible quantised value, rather than merely considering the current value and a single randomly-selected alternative.~\cite{MostChangedPixel} It is felt that this offers a better comparison between binary- and multi-level holograms as the randomness inherent in direct search algorithms is eliminated at the expense of higher computational requirements. 	
        
	\section{Effect on Convergence} \label{multi}
    
\begin{figure}[tbhp]
    	\centering
    	{\includegraphics[trim={0 0 0 0},width=1.0\linewidth,page=1]{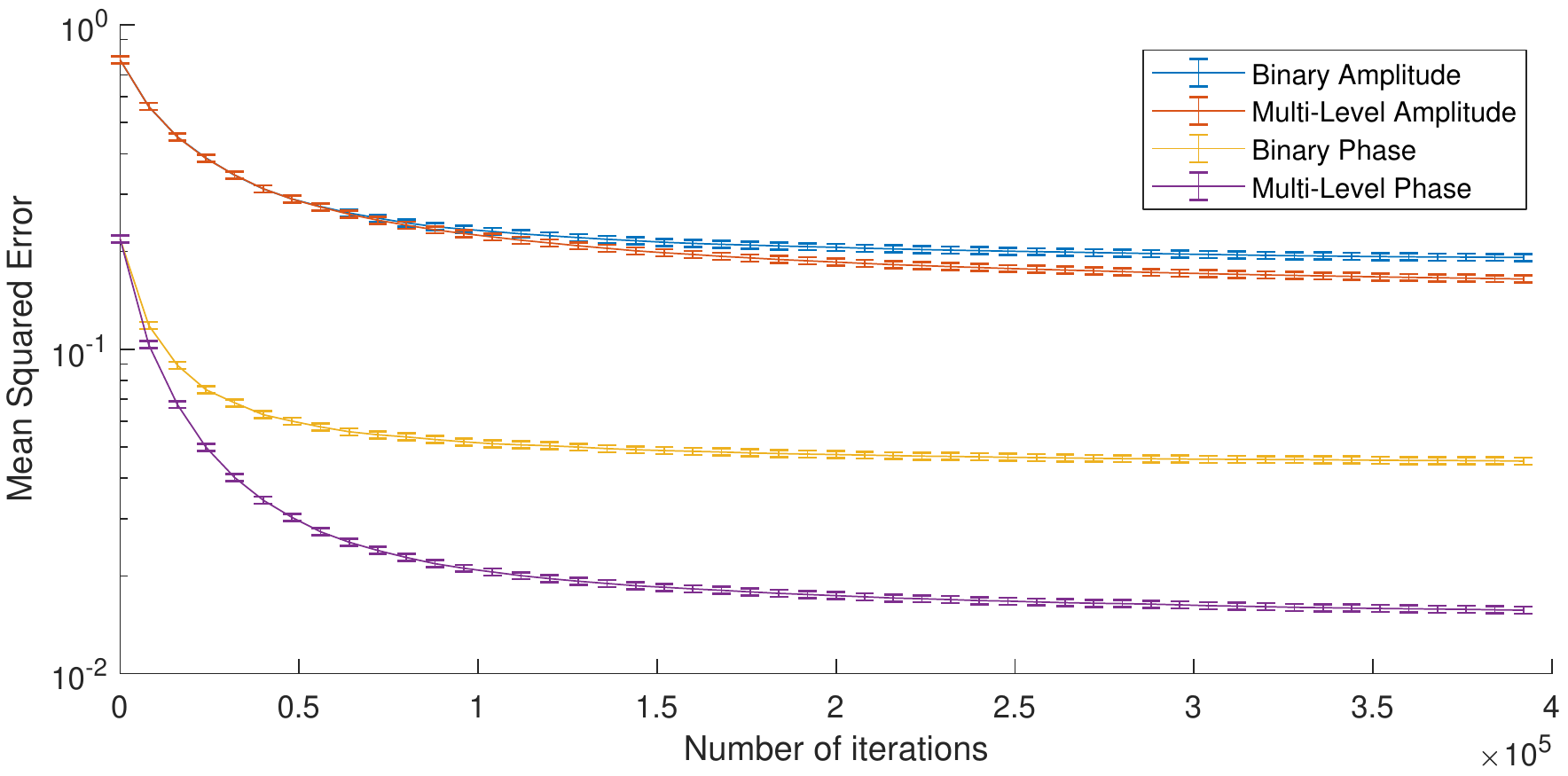}}
      	\caption{Comparison of convergence for multi-level and binary quantisation on phase-modulating and amplitude-modulating devices for the phase-insensitive replay field. Target is the \textit{Mandrill} test image shown in Figure~\ref{fig:MandrillRotSym} (left) with artificial rotational symmetry and a randomised phase profile. Values are taken as being the mean of 20 runs with independent random phase profiles with error bars showing two standard deviations. Error bars are shown for every 10,000th iteration to reduce visual clutter.}
      	\label{fig:convergence3}
    \end{figure}

    In order to confirm the above result in the context of whole image manipulation we perform $4\times 10^5$ iterations of the algorithm shown in Fig.~\ref{fig:MandrillRotSym} (right) for binary- and multi-level quantisation on phase- and amplitude-modulating SLMs. The results of this analysis are reported in Fig. \ref{fig:convergence3}, which shows the MSE convergence for the phase-insensitive replay field, and Table \ref{fig:stats}, which reports the final MSE obtained in each case. 
		
    \begin{table}[tbhp]
        \centering
        \caption{Algorithm convergence - final error as percentage of initial error}
        \begin{tabular}{c c c c}
            \toprule
            Quantisation & Amplitude           & Phase                            \\
            Levels  &   Modulation         &     Modulation                        \\
            \midrule
            Binary       &  $24.57\%$   & $22.76\%$    \\
            256-Level   &  $21.06\%$   & $5.522\%$    \\  
            \bottomrule
            
        \end{tabular}
        \label{fig:stats}
    \end{table}
            
\begin{figure}[tbhp]
        \centering
        {\includegraphics[trim={0 0 0 0},width=1.0\linewidth,page=1]{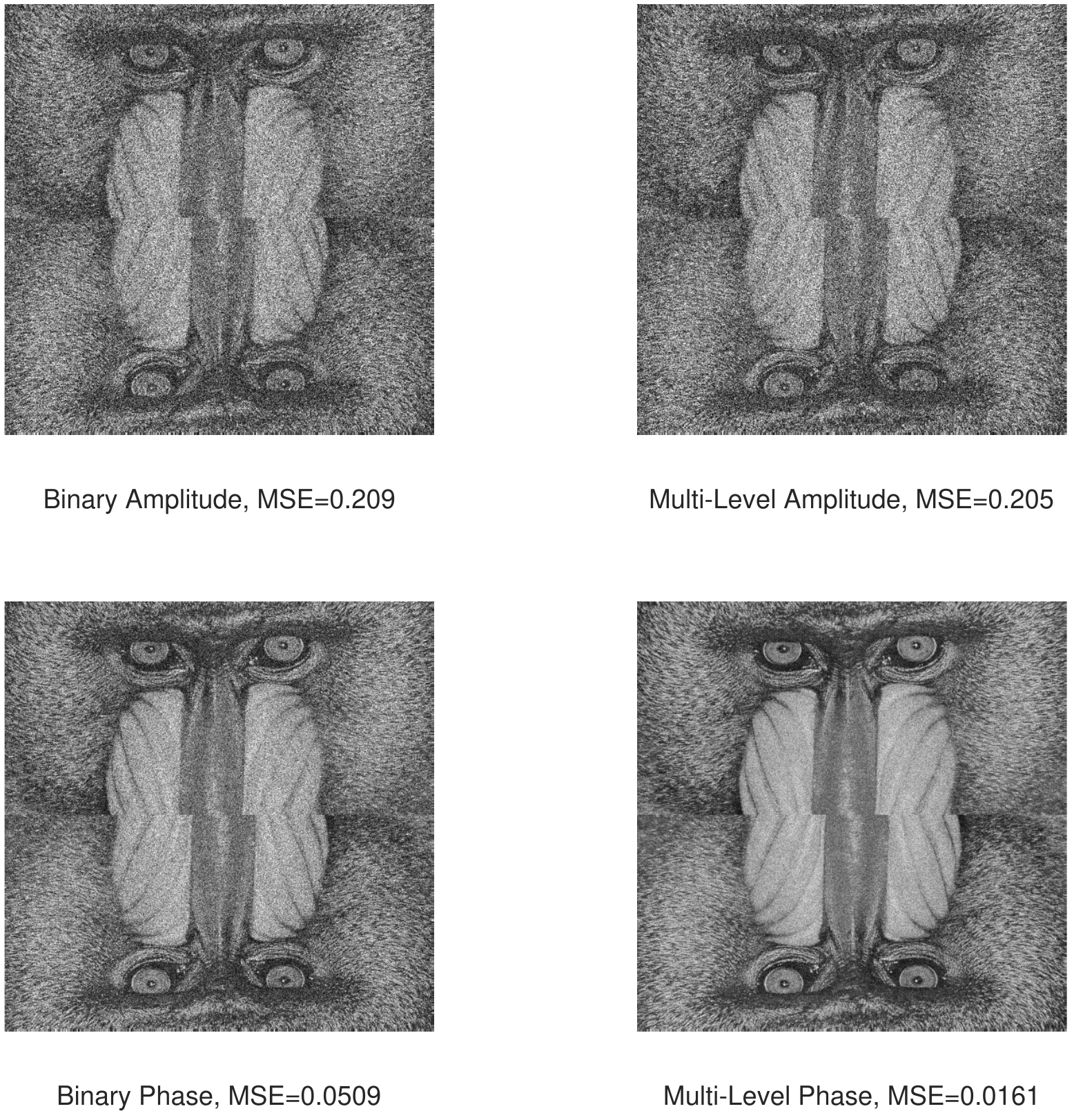}}
        \caption{Converged images corresponding to Fig.~\ref{fig:convergence3}. Target is the \textit{Mandrill} test image shown in Figure~\ref{fig:MandrillRotSym} (left) with artificial rotational symmetry and a randomised phase profile.}
        \label{fig:convergence4}
    \end{figure}

    It is observed that a $2^8$ level phase hologram offers a final error less than $25\%$ of the binary phase case. A $2^8$ level amplitude level hologram only offers a final error equal more than $85\%$ of the binary amplitude case. The final image qualities are shown in Fig. \ref{fig:convergence4}
    
    \section{Summary and Interpretation}
    
    This paper has discussed the relative merits of binary vs. multi-level quantisation in holography. We showed in Section~\ref{single} that on a pixel by pixel basis, the most desirable pixel phase rarely corresponds to that achievable on a phase device. By contrast, for amplitude devices we showed that on a pixel by pixel basis, the most desirable pixel value on a 256-level device is equal to the minimum or maximum value more than $85\%$ of the time. 
    
    To explore this further, in Section~\ref{multi} we took a modification of the DS algorithm to explore the convergence of the two algorithms over time. As expected, both amplitude and phase holograms converged to a hologram with non-zero reconstruction error. The surprising element is that while multi-phase holograms massively outperformed binary phase holograms, the same could not be said for multi-amplitude holograms which only slightly improved on the binary-amplitude case. 
    
    The authors believe that this result is due to the differences in modulation achievable between the regimes. A phase device does not change the amplitude of every spatial frequency, merely determining its location in the replay field. This leads to behaviour similar to that shown in Fig.~\ref{fig:changes} (left) where changing a single phase value has a sinusoidal impact on reconstruction error. An amplitude device, Fig.~\ref{fig:changes} (right), has a more complicated relationship as the angle of any one spatial frequency is fixed and it is the proportion that is variable. It can be imagined that for much of the time the actual preferred amplitude for a pixel would be either negative or greater than that achievable. In both cases, this would lead to the multi-level pixel being equal to the binary case.
    
    \section{Discussion}
    
    Our analysis here has offered evidence of an observation we feel is worthy of further exploration. It is worth listing some of the inherent limitations in the work, however:
    
    Firstly, the study has been entirely numerical in nature and does not take account of real-world issues such as lens aberration, non-flatness or speckle. Multi-level amplitude devices would offer a greater degree of flexibility when compensating for these issues.
    
    Secondly, the error metric used was MSE due to its simplicity of use. Display applications often use the structural simularity index (SSIM) as a metric of visual quality due to the greater range of quality issues highlighted. It has recently been argued that the distinction between SSIM and MSE is not as significant as previously thought \cite{Dosselmann2011} and future investigation should explore the effect of modulation on SSIM and other error metrics.
    
    Thirdly, the choice of algorithm deserves consideration. Many algorithms only function for a small number of different configurations. The search based algorithm used, Figure~\ref{fig:MandrillRotSym}, was chosen as it was equally applicable to every configuration considered \cite{DiLeonardo:07, Lee:79}. 
    
    Fourthly, local minima are an expected issue with search algorithms. The authors suggest that any effect of this will disproportionately effect binary devices and that the difference in convergence shown in Figure~\ref{fig:convergence3} is likely to overestimate the difference in \textit{best possible} hologram for binary and multi-level cases. While impossible to quantify exactly, the influence of local minima on the convergence graph can be estimated from the magnitude of the standard deviation in the independent runs.
    
    Fifthly, this analysis does not take account of more advanced optical configurations. A wide array of systems such as amplitude hologram encoding \cite{Kim:14} have been developed. These use amplitude SLMs in clever configurations to get a greater degree of control in the reconstruction. The analysis here applies only to a simple single-SLM hologram in either a 2f or far-field configuration and while it is expected that the results will be more widely applicable, this has not been explicitly researched. For example, phase holograms can emulate some of the behaviours of amplitude holograms \cite{Davis:99}.
    
    Sixthly, this work takes no specific account of the Fresnel regime, focussing exclusively on the Fraunhofer regime. While many of the assumptions made are equally applicable, this is worthy of further investigation.
    
    Seventhly, only a single test image, the \textit{Mandrill}, was considered. Independent tests with the \textit{Peppers} test image produced similar results but the effect of target amplitude profile on this result is an area requiring greater exploration.
    
    Eighthly, this analysis does not draw a conclusion on the relative merits of amplitude vs. phase holography, merely the effect of number of modulation levels. Other factors greatly influence that decision, for example phase SLMs are subject to phase compression and therefore require greater environmental control than their amplitude counterparts. Figure~\ref{fig:convergence3} fails to take this and many other effects into account.
    
    Ninthly, we do not draw a conclusion on the relative merits of amplitude vs. phase holography, merely the effect of number of modulation levels. Other factors greatly influence that decision, for example phase SLMs are subject to phase compression and therefore require greater environmental control than their amplitude counterparts. Figure~\ref{fig:convergence3} fails to take this and many other effects into account.
    
    Tenthly, this letter does not include a discussion of other concerns in holography other than that of final error. For example, diffraction efficiency considerations factor into amplitude hologram design \cite{Wyrowski:90}.
        
    Eleventhly, we have only considered the case of a phase insensitive replay fields.
    
    Twelthly, this work does not discuss any methods based on phase-only devices with a phase-shift larger than $2\pi$.
    
    Finally, the authors would be interested in this analysis's applicability to similar problems. For example, our approach might be applicable to a comparison of binary and continuous zone plates.
    
    \section{Conclusion}
	
	The impact of altering a single hologram pixel and of optimising an entire hologram have both been investigated for phase-only and amplitude-only holograms, with overall improvement metrics provided in Table.~\ref{fig:stats}. It has been shown that, for the Mandrill replay field of Fig. \ref{fig:MandrillRotSym} (left) generated using a variant of direct search, optimising the hologram for display on a 256-level phase SLM provides a significant improvement over optimising the hologram for display on a binary phase SLM. On the other hand, optimising the same hologram for display on a 256-level amplitude SLM only offers a marginal improvement over a binary amplitude SLM in both use cases. 
    
    A detailed discussion of the assumptions and limitations of this observation has been presented and it is hoped that this could help inform the decision making process during experimental design.	
    
    \section*{Acknowledgements}
    
    The authors would like to acknowledge Mr Ralf Mouthaan for his assistance in proofreading and correcting this manuscript.
    
    \section*{Funding}
    
    The authors would like to thank the Engineering and Physical Sciences Research Council (EP/L016567/1) for financial support during the period of this research.
    
    \section*{Disclosures}
    
    The authors declare no conflicts of interest.

    \bibliography{references}
    
\end{document}